# Transformation between elastic dipoles, quadrupoles, octupoles and hexadecapoles driven by surfactant self-assembly in nematic emulsion

Bohdan Senyuk[1], Ali Mozaffari[2], Kevin Crust[1], Rui Zhang[2,3], Juan J. de Pablo[2,4*], Ivan I. Smalyukh[1,5,6*]

[1]Department of Physics and Soft Materials Research Center, University of Colorado, Boulder, CO, USA.
[2]Pritzker School of Molecular Engineering, The University of Chicago, Chicago, IL, USA.
[3]Department of Physics, The Hong Kong University of Science and Technology, Clear Water Bay, Kowloon, Hong Kong SAR.
[4]Center for Molecular Engineering, Argonne National Laboratory, Lemont, IL, USA.
[5]Department of Electrical, Computer, and Energy Engineering, Materials Science and Engineering Program, University of Colorado, Boulder, CO, USA.
[6]Renewable and Sustainable Energy Institute, National Renewable Energy Laboratory and University of Colorado, Boulder, CO, USA.

*E-mail: depablo@uchicago.edu (J.J.D.P), ivan.smalyukh@colorado.edu (I.I.S.)

## Abstract

Emulsions comprising isotropic fluid drops within a nematic host are of interest for applications ranging from biodetection to smart windows, which rely on changes of molecular alignment structures around the drops in response to chemical, thermal, electric and other stimuli. We show that absorption or desorption of trace amounts of common surfactants can drive continuous transformations of elastic multipoles induced by the droplets within the uniformly aligned nematic host. Out-of-equilibrium dynamics of director structures emerge from a controlled self-assembly or desorption of different surfactants at the drop-nematic interfaces, with ensuing forward and reverse transformations between elastic dipoles, quadrupoles, octupoles and hexadecapoles. We characterize inter-transformations of droplet-induced surface and bulk defects, probe elastic pair interactions and discuss emergent prospects for fundamental science and applications of the reconfigurable nematic emulsions.

## Introduction

Emulsions are common soft matter systems associated with various foods and drinks, health and personal care products, bodily fluids essential for functioning of humans and animals, and many other (*1*). Nematic emulsions comprise nematic liquid crystals (LCs) as at least one of the fluid emulsion components (*2, 3*). They have recently attracted a great deal of applied and fundamental interest because of bringing about facile response of the LC to highly enrich the physical behavior of these systems (*4-18*). Nematic drops within isotropic fluids and nematic host materials with embedded isotropic droplets exhibit fascinating structures of defects and distortions in the LC molecular alignment field (*2-18*), which comply with the constraints of topological theorems (*13*). Such structures of molecular alignment within or around the droplets can serve as reporters of biological microbes or chemical sensors (*4, 5*), as well as can drive various forms of reconfigurable elasticity-mediated self-assembly (*2, 7, 11-13*). In emulsions comprising foreign fluid drops within an aligned nematic host, a variety of defect and structures have been reported, giving origins to particle-induced deformations of the far-field alignment that range from elastic dipoles (*2*) to quadrupoles (*7*, *16*), to hexadecapoles and to even higher-order multipoles (*17-19*). Some of these structures, which typically can be characterized by leading-order elastic multipoles, were found inter-transforming in response to applied electric fields or changes of temperature and other experimental conditions (*7*, *18*). Elasticity-mediated self-assembly of the droplets within the nematic host medium is highly dependent on the elastic multipolar nature of the director (**n**≡-**n**) distortions around the droplets, as well as on the structure and geometry of the corresponding particle-induced topological defects, and could be used in designing mesostructured composite materials (*2-20*). However, limitations of controlling the occurrence of different elastic multipoles and corresponding defect structures hinder the utility of LC emulsions in fundamental-science and technological applications.

Nematic emulsions are of great fundamental interest. How a sphere with well-defined boundary conditions interacts with a vector or director field is understood within the predictions of topological theorems, like the Hairy Ball theorem (*21*). However, topological theorems define the rules for the field-surface interactions only up to the topological characteristics, so that a number of topologically analogous but energetically similar structures can arise while having very different structures of particle-induced defects (*2*, *3*, *22*, *23*). In the context of LC colloids and emulsions, this manifests itself via various interesting topology-compliant director configurations appearing under different conditions, with a classic example being formation of either a hedgehog point defect or a Saturn ring disclination that both allow embedding a drop with perpendicular boundary conditions into a uniformly aligned nematic background (*2*, *8*, *13*, *16*). Transformations between some of these elastic multipole structures have been demonstrated, including when prompted by external fields (*7*, *18*, *24*, *25*), though these studies so far were limited to only few types of elastic multipoles. Mechanisms of transformations between various topologically admissible configurations of defects and elastic multipoles as transiently changing or stable structures are known for only limited types of surface boundary conditions (*7*), even though fascinating nematic emulsions and colloids with patchy surface anchoring have been demonstrated (*26-30*).

Here we report spontaneous transformations between elastic dipoles, quadrupoles, octupoles and hexadecapoles occurring in different forward and reverse sequences in response to absorption or desorption of surfactant molecules at the LC-isotropic fluid interfaces of droplets. While the elastic dipoles and quadrupoles are commonly studied (*2, 8, 13*) and high-order elastic multipoles have been reported to occur at very specific pre-designed boundary conditions on surfaces of solid particles (*31, 32*), our study reveals how these different multipoles can arise within an emergent hierarchical self-assembly process occurring at lengths scale from ~1 nm to tens of micrometers. Remarkably, these highly controlled transformations of elastic multipoles can occur in a forward and reverse order, in agreement with our computer simulations based on Landau-de-Gennes free energy minimization, when the system is driven to different equilibria corresponding to different types of behavior exhibited by surfactant molecules at the LC-isotropic fluid interfaces. We demonstrate how a systematic variation of patchy surface boundary conditions through controlling the surfactant molecule self-assembly triggers fusion, splitting and various other transformations of singular nematic defects at surfaces and in the bulk of the nematic host medium. We show that the ensuing formation and transformation of various elastic multipoles can progress under out-of-equilibrium conditions, being driven by absorption and self-assembly of surfactant molecules, or the reverse process. The newly observed elastic multipoles of different orders and signs, each with a characteristic configuration of droplet-induced topological defects, may enable new forms of colloidal self-assembly (*13*), nanoparticle templating by defects (*33*) and biodetection (*4*), whereas their fascinating inter-transformation sequences can be used to design programmable reconfigurability of the mesostructured nematic emulsions.

**Results**

### Droplet-induced elastic multipoles

Far away from droplets and both bulk or surface defects induced around the isotropic droplet surfaces (at distances much larger than the droplet dimensions), the director distortions with respect to the uniform far-field background can be represented as a summation over the elastic multipoles (*31, 32*):

$$n_\mu(r, \theta, \varphi) = \sum_{l=0}^{\infty} \sum_{m=-l}^{+l} q_{lm}^\mu \frac{r_0^{l+1}}{r^{l+1}} Y_l^m(\theta, \varphi), \qquad (1)$$

where $\theta$ and $\varphi$ are respectively polar and azimuthal angles, $Y_l^m(\theta, \varphi)$ are spherical harmonics, $q_{lm}^\mu$ are dimensionless elastic spherical multipole coefficients, $l$ determines the order of a multipole as the $2^l$-th pole, $-l \leq m \leq l$, and $r_0$ is a radius of the droplet. One can immediately notice the analogy between the multipole expansion of the droplet-induced director distortions in the far-field and the corresponding expansion of electrostatic charge distributions (*31, 32, 34*). This analogy stems from the fact that the nematic LC elastic free energy density within the one-elastic-constant approximation reads

$$f_E = \frac{1}{2} K \sum_{\mu=x,y} (\nabla n_\mu)^2, \qquad (2)$$

where $K$ is the average Frank elastic constant and $n_\mu$ ($\mu = x, y$) are director components perpendicular to the far-field director $\mathbf{n}_0$ (along the $z$-direction). We note that the use of this one-constant approximation in this case is crucial to the electrostatics analogy and is justified by the fact that the different elastic constants often have comparable values (*34*). The structures of elastic colloidal multipoles minimize this free energy density that gives Euler-Lagrange equations in the form of the Laplace equations, bringing about the electrostatic analogy (*34*):

$$\nabla^2 n_\mu = 0. \qquad (3)$$

Multipole moments $q_{lm}^\mu$ can be determined with the following integral

$$q_{lm}^\mu = \int_0^{2\pi} \int_0^\pi n_\mu(r, \theta, \varphi) \frac{r^{l+1}}{r_0^{l+1}} Y_l^{m*}(\theta, \varphi) d\theta d\varphi, \qquad (4)$$

assuming $n_x, n_y \ll 1$ and $n_z \approx 1$ far from the particle surface. While elastic colloidal dipoles ($l=1$) and quadrupoles ($l=2$) in emulsions and solid particle dispersions have been widely studied for over two decades (*2, 13*), higher-order multipoles like hexadecapoles ($l=4$) were observed only recently and typically require rather special shape or surface functionalization to exist (*17, 18, 31, 32*). While symmetry-allowed higher-order multipoles are always present in the above expansion, the question often is if they need to be considered, which depends on relative strengths of different multipole moments (*13*). Interestingly, our findings below show how self-assembly and desorption of surfactant molecules at LC-glycerol interfaces drive transformations between highly diverse director structures that, up to leading order, can be described as multipoles with $l$=1, 2, 3, 4.

**Topological transformations from boojums quadrupole to hedgehog dipole**
Micron-sized spherical droplets of isotropic fluid in a nematic LC were obtained by mixing a small amount of glycerol (fig. S1A) (~1-5 vol. %) with a nematic LC and vigorous agitation via a combination of flicking and sonication (see Materials and Methods). These droplets maintain their spherical shape when mixed with the LC because of the high interfacial tension. The emulsions are infiltrated into glass cells of about 30-60 µm gap thickness, much larger than the diameter of the droplets. To satisfy the boundary conditions at the droplets surface, the LC director field **n**(**r**) deviates from the far-field director **n**$_0$ defined by boundary conditions at the confining glass plates and forms characteristic patterns with accompanying topological defects. These different director perturbations are characterized by the ensuing elastic colloidal multipoles (*13*, *31*, *32*, *34*). When pure glycerol is used to form droplets dispersed in 5CB or ZLI-2806 nematic LCs, planar azimuthally degenerate boundary conditions are naturally promoted at the LC-glycerol interface (*3*). This results in a quadrupolar configuration of **n**(**r**) around the droplet (Fig. 1Aa) with two surface point defects (called "boojums") at its poles on diametrically opposite sides of the droplet along **n**$_0$ (Fig. 1Ba and Ca). Such droplets form an elastic quadrupole with $l$=2 and $m$=+1 (Fig. 1Aa-Da) (*7*, *31*), which we will refer to as a boojum quadrupole. Boojum quadrupoles obtained with a pure glycerol are stable over long periods of time without changing the **n**(**r**)-configuration. However, adding a trace amount of surfactant to glycerol before mixing with the LC alters boundary conditions and changes this droplet-induced elastic multipole behavior.

Surfactant sodium dodecyl sulfate (SDS) (fig. S1B) uniformly dissolves within glycerol at studied small concentrations. At SDS concentrations >0.1 wt. %, homeotropic boundary conditions are induced at droplet surfaces, giving the origins to a dipolar **n**(**r**)-configuration with a bulk point defect (Fig. 1Ah). For SDS at <0.1 wt. %, the droplets initially appear as boojum quadrupoles with tangential boundary conditions (Fig. 1Aa, Ba and Ca) right after the preparation, but **n**(**r**)-configurations and singular defects change with time. First, two boojums at the poles open into the quarter-integer surface-bound defect loops encircling the surface of the droplet at its poles. Later, these surface defect loops slowly drift away from the poles (Fig. 1Aa-Ad, Movie S1) and, upon reaching the equator, combine to form a single half-integer bulk defect loop called "Saturn ring," as well as the corresponding quadrupolar **n**(**r**)-configuration with $l$=2 and $m$=-1 (*16*, *31*) (Fig. 1Ae). Then, the Saturn ring starts moving towards one of the poles of the droplet (Fig. 1Af and Ag), eventually transforming into a bulk point defect of elementary hedgehog charge; the corresponding dipolar **n**(**r**)-configuration (Fig. 1Ah) has $l$=1 and $m$=+1. Polarizing and brightfield microscopy textures (Figs. 1A and 1B, respectively) reveal details of these transformations, with the corresponding schematics and numerical visualizations of elastic multipoles shown in Figs. 1C and 1D. It is interesting to note that, during the first phase of the transformations (Fig. 1Ab-Ad), when two quarter-integer surface defect loops are moving towards the equator, the polarizing micrographs feature eight bright lobes around the droplet (fig. S2A), as in the elastic hexadecapoles with conically degenerate surface anchoring (fig. S2E) studied recently (*17-19*). These experimentally observed topological transformations of elastic multipoles are caused by tangential-to-homeotropic changes of patchy surface anchoring boundary conditions at the LC-glycerol interface (Fig. 2A-C). SDS molecules are initially evenly dispersed within the volume of the glycerol droplet (Fig. 2A), but with time they diffuse to and self-organize at the LC-glycerol interface and cause the surface anchoring transition. Nucleation of surfactant monolayer islands with homeotropic anchoring typically starts at the poles of the droplet (Fig. 2B) because the ensuing local change of boundary conditions minimizes the corresponding elastic energy. Self-assembled surfactant islands with homeotropic boundary conditions at the poles open the boojums into quarter-integer surface defect loops and, with time, transform elastic quadrupoles into hexadecapoles with mixed anchoring (Figs. 1Bb and 2B), different from hexadecapoles studied previously (*17, 1*8). Increasing the surfactant density leads to the growth of these patches with homeotropic boundary conditions, expanding dynamics of the surface defect loops that then move toward the equator (Fig. 1Bb-Bd). As the two patches with homeotropic boundary conditions meet at the equator and two quarter-integer surface disclination loops merge to form a half-integer bulk disclination loop, the elastic hexadecapole with mixed anchoring transforms into an elastic quadrupole with a Saturn ring (Figs. 1Ae and 2C). The transformation from a Saturn ring quadrupole to an elastic dipole is nearly twice faster than the transformation from a boojum quadrupole to a Saturn ring quadrupole (Fig. 2D). Figure 3 shows the corresponding results of numerical computer simulations, where boundary conditions are continuously changed from tangential to homeotropic starting from the poles, mimicking the experimental observations (Materials and Methods). Computer-simulated polarizing optical microscopy textures of transformations of the director structure around the droplet and corresponding elastic multipoles (Fig. 3, Movies S2) are in excellent agreement with the experimental observations.

By controlling the amount of the added surfactant, one can stop topological transformations at different desired stages with a specific director configuration around the droplet and a resulting elastic multipole. Although topological transformations from a boojum quadrupole to an elastic dipole are rather interesting, even more exciting are the intermediate states that correspond to the observed elastic hexadecapoles (Figs. 3Ab,c and 4) and octupoles (Figs. 3Db-d and 5) formed by droplets with mixed patchy tangential-homeotropic surface anchoring. The growth of self-assembled surfactant islands is an out-of-equilibrium process that enriches accessible structures during transformation by prompting the appearance of states that cannot be found during conventional switching transitions, say when induced by electric fields (*7, 25*). As we discuss in detail below, the observed transformations of elastic multipoles conserve topological charges of the induced defects like boojums, hedgehogs and Saturn rings while they all compensate the effective topological charge due to uniform or mixed boundary conditions on the spherical droplet's surface (*7, 13, 21, 34*).

**Elastic hexadecapoles with mixed patchy anchoring**
The configuration of $\mathbf{n}(\mathbf{r})$ around the droplet intermediate between a boojum quadrupole and a Saturn ring quadrupole is hexadecapolar (Figs. 3Ab,c and 4, fig. S2). A polarizing optical micrograph (Fig. 4A) shows eight bright lobes around the perimeter of the droplet separated by narrow dark regions within which $\mathbf{n}(\mathbf{r})$ is parallel to polarizer or analyzer. Director field deviates away from $\mathbf{n}_0$ in the regions of bright lobes. Additionally, insertion of the phase retardation plate into the polarizing microscope's optical path reveals that the tilt of $\mathbf{n}(\mathbf{r})$ with respect to $\mathbf{n}_0$ is different within adjacent bright lobes (Fig. 4B) and switches between clockwise (blue color in the texture) and counterclockwise (yellow color in the texture) directions eight times along the perimeter. These textural features and $\mathbf{n}(\mathbf{r})$ are similar to the ones exhibited by an elastic hexadecapole (fig. S2E-H) previously found forming around colloidal particles with conically degenerate surface anchoring (*17-19*). However, the $\mathbf{n}(\mathbf{r})$-structure and defects in the hexadecapoles with tilted anchoring are different from that in hexadecapoles with mixed anchoring observed here. First, brightfield textures of hexadecapoles with mixed anchoring (Fig. 4C) reveal the presence of two quarter-integer surface defect loops, which separate areas with tangential and normal boundary conditions (Figs. 1Ab,c and 3Ab,c) whereas hexadecapoles with conically degenerate anchoring have two boojums at the poles and one defect loop at the equator (compare Fig. 4A-D and fig. S2E-G). Second, the hexadecapolar structure observed in the current study has four bright lobes corresponding to the areas with tangential boundary conditions and four bright lobes close to the poles correspond to the areas with homeotropic boundary conditions of the droplet. Experimental textures (Fig. 4A-C) reveal the detailed **n(r)** around the droplet (Fig. 4D), which is in good agreement with numerical modeling (Fig. 3Ab or Ac) (*17-19*, *31*). A color map of the projection $n_x$ of **n(r)** onto the *x*-axis orthogonal to **n**$_0$ highlights positive, near zero and negative $n_x$ (Figs. 3Bb,c, 4E) around the droplet and is also consistent with a hexadecapolar **n(r)**-configuration (*17-19*, *31*, *32*). Mapping of $n_x$ at some distance from the spherical surface reveals that the tilting of **n(r)** with respect to **n**$_0$ in the hexadecapole with mixed anchoring has an opposite sequence as compared to that in the hexadecapole with conic anchoring (compare Fig. 4E, fig. S2B,D and fig. S2F,H). Thus, both hexadecapoles have the same *l*=4 leading multipole order but differ in that we have *m*=+1 for the hexadecapoles with conic anchoring and *m*=-1 for the hexadecapoles with mixed anchoring (*31*).

Sharing of the far-field deformations of the director between different elastic multipoles give rise to elasticity-mediated interactions. At center-to-center distances much larger than the droplet size, the colloidal pair-interaction potential, derived from an electrostatic analogy of the far-field distortions (*17*), reads

$$U_{int} = 4\pi K \sum_{l,l'=1}^{N} a_l a'_{l'} (-1)^{l'} \frac{(l+l')!}{R^{l+l'+1}} P_{l+l'}(\cos\theta), \tag{5}$$

where *R* is a distance between the centers of two interacting droplets, $a_l = b_l r_0^{l+1}$ is an elastic multipole moment of the *l*th order, $P_{l+l'}(\cos\theta)$ are the Legendre polynomials and $\theta$ is an angle between **n**$_0$ and the center-to-center droplet separation vector **R**. Even though Eq. (5) is accurate only at large distances, it gives qualitative insights into the pair interaction potential behavior even at distances corresponding to only few particle sizes, though singular defects and near-field effects start playing important roles when *R* is becoming comparable to the droplet diameter. The $\theta$-dependent elastic pair interactions of hexadecapoles have eight zones of attraction separated by eight zones of repulsion (*17-19*), which correlate with eight bright lobes in polarizing optical micrographs. Two droplets attract if zones of the same color (Fig. 4B) in the polarizing textures with a phase retardation plate radially face each other. Pair interactions of hexadecapoles with mixed anchoring depend on relative areas of patches with dissimilar anchoring, with quadrupolar interactions dominating when their size is very different (Fig. 3Ab, Bb). The quadrupolar moment is at minimum and hexadecapole moment is maximized when the size of the bright (Fig. 4A) or blue and yellow (Fig. 4B) lobes along the droplet perimeter in polarizing textures are comparable (see also Fig. 3Ac, Bc). For example, the ratio between the hexadecapolar ($b_4$) and quadrupolar ($b_2$) moments obtained from fitting (Fig. 4F) is $b_4/b_2 \approx 1.93$ at $\theta \approx 38°$. While the far-field director structure and interactions of such droplets are consistent with their leading-order hexadecapole nature, it is important to note that the near-field effects and proximity of singular defects alter this behavior at small distances.

**Elastic octupoles with mixed anchoring**
Another interesting **n(r)**-configuration around the droplet and corresponding elastic multipole can be obtained when one of the boojums opens into the growing surface defect loop but the other remains at the pole (Fig. 5, Movies S3). In our experiments, because of the associated nucleation and growth process, such asymmetric evolution of structures often happens spontaneously, but this process can be controlled if a small foreign inclusion is trapped by one of the boojums but not the other, thus resulting in different energetic barriers for the surfactant island self-assembly at the two poles. In this case, polarizing micrographs (Fig. 5A) show only six bright lobes around the droplet perimeter. Tilt of **n**(**r**) with respect to **n**$_0$ is different within adjacent bright lobes (Fig. 5D) and switches between clockwise (blue color in the texture) and counterclockwise (yellow color in the texture) directions with respect to **n**$_0$, like in the hexadecapole with mixed anchoring, but only six times. The brightfield micrograph (Fig. 5B) reveals presence of one quarter-integer surface disclination in the lower droplet's hemisphere and the boojum defect at the top pole. The experimentally reconstructed **n**(**r**) around such a droplet shown in Fig. 5C is in a good agreement with numerical calculations (Fig. 3Db-d). A color map of the projection $n_x$ of **n**(**r**) onto the *x*-axis orthogonal to **n**$_0$ (Fig. 5E) closely resembles the electrostatic octupole (*13*, *31*), though here its elastic octupolar counterpart is formed by induced director distortions around a colloidal inclusion undergoing anisotropic diffusion with different coefficients ($D_\parallel/D_\perp$=1.35) measured parallel and perpendicular to **n**$_0$ (fig. S3G,H) (*35*, *36*).

Two elastic octupoles attract (Fig. 5F) if zones of the same color in the polarizing micrographs with a retardation plate (Fig. 5D) align radially more or less against each other, so that the angular dependence of their elastic pair interactions has six zones of attraction separated by six zones of repulsion (Fig. 5H and inset of Fig. 5G). This behavior correlates with the angular diagram of elastic interactions of two anti-parallel octupoles (Fig. 5H and inset of Fig. 5G) calculated using elastic moments ($b_1,b_3,b_5$)=(0.182, 0.499, -0.005), obtained from fitting the potential (Fig. 5G) extracted from the experimental attraction between two droplets (Fig. 5F). We note that the dependence described by Eq. (5) is strictly speaking expected to hold only at large distances, but it provides qualitative insights even at R corresponding to several droplet sizes; here, like for hexadecapoles discussed above, singular defects and near-field effects modify the short-range elastic interactions when droplets closely approach each other. The angular dependence of interactions (Fig. 5H) will change for parallel octupoles: zones of attraction will become zones of repulsion and vise versa. The elastic interactions potential of octupoles amounts to hundreds of $k_\text{B}T$ (Fig. 5G) and is an order of magnitude weaker than that for elastic dipoles and quadrupoles formed by colloidal inclusions of comparable dimensions (*13*). The strength of the octupolar moment depends on the ratio between the area of patches with homeotropic boundary conditions and the area of patches with tangential boundary conditions between the equator and a surface defect (Fig. 5A-E and fig. 3D-F). The quadrupolar moment is dominant when the area with homeotropic boundary conditions is small and a dipolar moment increases when the surface defect line is close to the equator and homeotropic boundary conditions cover nearly half of the droplet. The octupolar moment is at maximum ($b_3/b_1 \approx 2.7$) when the quarter-integer surface defect line is somewhere in the midway between the pole and the equator (Fig. 5D,F).

The elastic octupole can transform into the elastic dipole if there are enough SDS molecules to cover the whole LC-glycerol interface (Fig. 5I) and change boundary conditions to homeotropic all over the droplet's surface. In this case, the quarter-integer surface disclination loop formed by opening the boojum at the bottom pole travels all the way from the bottom pole to the top pole, where it merges with the boojum surface point defect, which, in turn, produces a bulk hedgehog point defect.

**Topological transformations from hedgehog dipole to boojum quadrupole**
Topological transformations in the opposite direction, from an elastic dipole to a boojum quadrupole, also occur within the inverse LC emulsions. To demonstrate them, as well as to show that the studied transformations are rather universal for LCs and not specific to a certain nematic compound or a mixture, we used a nematic 5CB as a host medium for the glycerol droplets with trace amounts of water, different from a multi-component nematic mixture studied above. An alignment agent N,N-dimethyl-N-octadecyl-3-aminopropyl-trimethoxysilyl chloride (DMOAP) (fig. S1C) was then used as a surfactant mixed with glycerol at <0.1wt.%, sufficient to initially cover the entire droplet's surface by the surfactant monolayer. Figure 6A shows the sequence of brightfield textures during this transition. Polarizing micrographs reveal that a hyperbolic hedgehog point defect at one of the poles and the ensuing dipolar $\mathbf{n}(\mathbf{r})$-configuration right after the sample preparation (Fig. 6Aa), when strong homeotropic boundary conditions (Fig. 6Ba) are defined by DMOAP molecules at the interface. With time, however, this dipolar configuration becomes unstable and the hedgehog opens into the half-integer disclination loop, the Saturn ring, which then expands and moves towards the equator (Fig. 6Ab, Movie S4). This transformation of the hedgehog into the Saturn ring configuration can be explained by weakening of the homeotropic surface anchoring which might be caused by a slow dissociation of DMOAP molecules from the interface (*37*). While full understanding of the nature of this anchoring weakening at the interface will require separate detailed studies, a possible reason is that the trace amounts of water within glycerol cause DMOAP molecules to dimerize/polymerize and be less effective in inducing the homeotropic interfacial boundary conditions for the director and even eventually to dissociate from the interface. The change of the surface anchoring strength alters the interplay between surface and bulk elastic free energy contribution, causing the out-of-equilibrium structural transformations that allow the system to reduce the overall free energy via transformation to the quadrupolar structure (*24*, *25*, *38*). After the Saturn ring of a half-integer disclination reaches the equator, it splits into two quarter-integer surface disclinations moving towards the droplet poles (Fig. 6Ac,d, Movie S4). The Saturn ring splits into two disclinations when boundary conditions at the equator locally change from homeotropic to tangential, being a reverse process of what we described above during the surfactant self-assembly causing the growth of homeotropic anchoring patches. The two quarter-integer surface disclinations separate regions of dissimilar, tangential and homeotropic, boundary conditions and move towards poles as an area with tangential boundary conditions continuously grows out from the near-equator region. The area with tangential boundary conditions nucleates at the equator because a change of boundary conditions from homeotropic to tangential decreases director deformations [$\mathbf{n}(\mathbf{r})$ becomes parallel to $\mathbf{n}_0$] at the equator and corresponding elastic energy. The loops of quarter-integer surface disclinations eventually contract into two boojums (Fig. 6Ae, 6C, Movie S4) after they reach the poles of the droplet. These transformations of elastic multipoles from an elastic dipole to a boojum quadrupole undergoes with conservation of topological charges and becomes possible due to the change of boundary conditions at the LC-glycerol interface from homeotropic to tangential. The patchy surface anchoring transition likely happens due to hydrolysis of silane groups and polymerization of individual DMOAP molecules initially located at the LC-glycerol interface but eventually desorbing from it. This patchy out-of-equilibrium anchoring transition drives transformation of multipolar structures that is remarkably similar (but in opposite direction) to that of a boojum quadrupole to a hedgehog dipole transformation discussed above. Numerical modeling once again reproduces the entire transformation sequence while additionally revealing the final stage of transformation when boojums inherit the half-ring handle-like geometry (Fig. 6C, Movie S5), similar to that previously studied for solid colloidal particles with different genus (*22, 39*).

**Fusion, splitting and other transformations of singular defects**

Structures of dipolar and quadrupolar director distortions, with accompanying hedgehog, Saturn ring and boojum defects, are widely studied and well understood (Fig. 1Ca, Ce and Ch) (*2, 3, 13, 22, 23*). In all these cases, the droplet-induced defects appear to match well defined tangential or homeotropic boundary conditions at the LC-droplet interface to the uniform far-field alignment. A spherical glycerol droplet has a genus equal to zero and Euler characteristic $\chi_{Euler}=2$. Consistent with topological theorems (*2, 3, 13, 22, 23*), the hedgehog charge of the three-dimensional topological defects $\pi_2(\mathbb{S}^2/\mathbb{Z}_2)=\mathbb{Z}$ induced in the nematic LC bulk (the hyperbolic hedgehog point defect and the equivalent to it Saturn ring of $\pi_1(\mathbb{S}^2/\mathbb{Z}_2)=\mathbb{Z}_2$ disclination) compensates the hedgehog charge of surface boundary conditions at the drop surface is equal $\pm\chi_{Euler}/2=\pm1$; the sign of the induced defect can be determined upon decorating the non-polar director field with a unit vector field along one of the two anti-parallel directions (*23, 40*). The net winding number of $\pi_1(\mathbb{S}^1)$ LC defects in the two-dimensional interfacial director field at the LC-glycerol interface is equal to $\chi_{Euler}=2$, with the two boojums each having a winding of unity (Figs. 1Ca and 7). When the core of a boojum splits into a handle-like disclination loop (Fig. 7A), the interfacial director field than contains four defects with half-integer winding numbers. The intermediate structures of topological defects are subject to the same topological theorem constraints, albeit now for the mixed tangential-homeotropic patchy boundary conditions (*2, 3, 13, 22, 23*). An important feature of the intermediate structures observed during transformations is the presence of fractional surface defect lines. Unlike in the bulk, the winding number of disclinations is not constrained to be an integer multiplied by ±1/2, so that fractional disclinations are allowed too (*41*). The ones observed in our study are quarter-integer disclinations with the winding number of -1/4 (Figs. 1C, 4D, 5C and 6B), where this winding number is found by locally circumnavigating the surface defect line and measuring the director rotation within the LC in the proximity of this defect, divided by $2\pi$. While past studies demonstrated junctions of such quarter-integer surface defect lines and the bulk half-integer defect lines (*41*), our current study shows how two quarter-integer surface disclinations can combine to form a single loop of the bulk half-integer disclination $\pi_1(\mathbb{S}^2/\mathbb{Z}_2)=\mathbb{Z}_2$ (Fig. 1C), as well as that a reverse process can take place (Fig. 6B) while preserving conservation of topological characteristics. The forward-directed transformations of elastic multipoles involve fusion of quarter-integer surface disclinations into a single half-integer bulk defect loop (Fig. 1Cc-e) and its splitting into two quarter-integer surface disclination loops is observed during the reverse process (Fig. 6Bb-d).

Another interesting aspect related to topological defects is associated with the nature of nucleation of surfactant monolayer islands (Fig. 7A-N). Consistently with experiments, the numerical modeling shows that the free energy minimization of the LC host medium favors nucleation of surfactant islands at the droplet poles, at the location of point-like or split-core handle-like surface boojums (Fig. 7O), though even the islands that nucleate away from poles are strongly attracted to the poles. The diverse set of possible structures in our system includes boojums with point-like cores and handle-like split cores, quarter-integer disclination loops around surfactant islands replacing the boojums, or co-existing and interacting with them (Fig. 7 A-N). Interestingly, a local minimum and a non-monotonic behavior of free energy versus displacement is observed for the islands of surfactant with homeotropic patch of anchoring nucleated away from the poles; this behavior arises as a result of interaction and fusion of distinct singular cores of a boojum defect and a disclination loop surrounding the island. While the structures of cores of defects with locally reduced scalar order parameter look more complex when surfactant islands nucleate away from the top/bottom poles (Fig. 7D, E, I, M, N and inset of O), their net charge in the interfacial director field still adds to $\chi_{Euler}=2$, with a unity winding number per top/bottom hemisphere. Well-defined defect structures are important for controlling spatial localization of various nanoparticles and even self-assembled molecular structures (*13*), which makes our finding particularly interesting because the observed transformations of defect structures could be potentially utilized for reconfiguration of defect-entrapped nanostructures.

## Discussion

Our results indicate that self-assembly of surfactant molecules can drive transformation of defects and director structures surrounding isotropic liquid droplets within LC emulsions formed from both single-compound LCs like 5CB and various nematic mixtures. Our study can be extended to explore the differences between such transformations in emulsions with single-and multi-component LCs, where various forms of interfacial phase separation could play a role in the latter case. Our observations for large droplets of size $r_0 >> K/W$ are consistent with the regime of strong polar anchoring imposed by patches of the droplet surface with or without surfactant monolayers. Indeed, the polar anchoring surface boundary conditions for micrometer-sized drops and $K \sim 10^{-11}$ N and $W \sim 10^{-5}$ J/m$^2$ can be considered as strong because the minimization of the total free energy mainly occurs through relaxation of the director structure within the LC bulk while keeping the director at LC-droplet interfaces fixed along the spatially varying easy orientation determined by the patchy distribution of surfactant molecules at the LC-droplet interface. One can envisage even richer behavior occurring as the droplet size would be reduced first to the regime of soft boundary conditions, where surface and bulk terms could be comparable and compete, and then to the regime of small drops when the surface energy would be relaxed while keeping bulk director distortions at minimum. The latter regime could be analogous to what was already observed for LC nanodroplets (*42*), so that the LC surrounding could drive patterning of surfactants within the droplet surfaces without elastic director distortions around the droplet. Isotropic fluid droplets, surfactant monolayers and the surrounding LC could all be polymerized by fixing respective configurations through crosslinking, as well as could be decorated with various nanoparticles, potentially offering new routes for hierarchical assembly of structured materials. When using microfluidics, droplet dimensions and amounts of accessible surfactants could be controlled rather precisely, potentially allowing for fabrication of emulsions with monodisperse droplet sizes, surfactant coverage and types of elastic multipoles induced by the droplets within the emulsions (*13*). This could allow for probing elasticity-mediated colloidal self-assembly due to high-order elastic multipoles and formation of various mesostructured composite materials enabled by self-assembly of reconfigurable elastic colloidal multipoles. Nematic emulsions with out-of-equilibrium structural transformations could be also engineered for various biodetection applications (*4*, *14*, *15*), where different elastic multipoles and defect structures could potentially serve as reporters of biological molecules and microbes at the LC-fluid interfaces (note that water could be used instead of glycerol).

## Materials and Methods

### Samples preparation and imaging techniques

Two different LCs were used as hosts for the glycerol (Sigma-Aldrich) (fig. S1A) droplets: 4-cyano-4´-pentylbiphenyl (5CB from Frinton Laboratories, Inc.) and ZLI-2806 (EMD Performance Materials Corp.). A small amount (<0.1 wt.%) of surfactant had been mixed with glycerol to alter a surface anchoring for LC molecules at the interface between LC and glycerol droplets. We mixed glycerol with a sodium dodecyl sulfate (SDS) surfactant (fig. S1B) for droplets in ZLI-2806 and an alignment agent N,N-dimethyl-N-octadecyl-3-aminopropyl-trimethoxysilyl chloride (DMOAP) (fig. S1C) for droplets in 5CB. To obtain micron-sized spherical droplets in a nematic host, the glycerol mixture was added to the LC, and then vigorously stirred using a combination of flicking and sonication. The ensuing diluted LC emulsions with spherical droplets ($r_0 \approx$ 1-10 μm) suspended evenly in a LC host were infiltrated into ~30-60-μm-thick cells made of two glass plates separated by glass spacers and sealed with a UV-curable glue. A polyimide PI2555 (HD Microsystems) was spin-coated on the glass plates, baked at 270 °C and uni-directionally rubbed with a velvet cloth to promote a tangential alignment of the LC in the cell. Experimental samples were stable over at least several days and LC structures around droplets were studied by means of brightfield and polarized light optical microscopy with an inverted Olympus IX81 microscope and a ×100 (NA=1.42) oil objective. Optical trapping and manipulations of the droplets were performed by holographic optical tweezers operating at 1064 nm (*43*). Microscopy images acquisition and analysis were performed using a CCD camera (Flea, PointGrey) and ImageJ software, respectively.

**Details of computer simulations**

Numerical continuum simulations were performed to study dynamics of nematic tensorial order parameter **Q**. For uniaxial systems, the nematic order parameter is written in the form of $\mathbf{Q} = S(\mathbf{nn} - \mathbf{I}/3)$ where unit vector **n** is the nematic director field and $S$, the largest eigen value of **Q**, quantifies the degree of uniaxial alignment. Within nematodynamics framework, the evolution of nematic tensorial order parameter follows the Beris-Edwards formalism (*44*):

$$\left(\frac{\partial}{\partial t} + \boldsymbol{u} \cdot \boldsymbol{\nabla}\right)\mathbf{Q} - \mathbf{S} = \Gamma \mathbf{H}, \qquad (6)$$

where **S** is generalized advective term which accounts for the response of the nematic order parameter to the symmetric and antisymmetric parts of velocity gradient. It is worth noting that the experimental observations reveal the quasi-static nature of nematic director evolution which this slow dynamic justifies simplifying the hydrodynamic equations and neglecting the reorientation of the director with backflow effects. As such, Eq. 6, reduces to the Ginzburg-Landau equation:

$$\frac{\partial \mathbf{Q}}{\partial t} = \Gamma \mathbf{H}. \qquad (7)$$

The molecular field

$$\mathbf{H} = -\left(\frac{\delta \mathcal{F}_{LdG}}{\delta \mathbf{Q}} - \frac{\mathbf{I}}{3}\mathrm{Tr}\frac{\delta \mathcal{F}_{LdG}}{\delta \mathbf{Q}}\right), \qquad (8)$$

embodies the relaxational dynamics of the nematic, which drives the system toward the minimum energy configuration. The free energy of the system is defined using a phenomenological Landau-de Gennes theory which reads

$$\mathcal{F}_{LdG} = \int_V f_{LdG} dV. \qquad (9)$$

The free energy density, $f_{LdG}$, is sum of bulk (describing isotropic/nematic phase transition) and elastic energies (accounting for penalizing nematic distortion from the uniform configuration) given by:

$$f_{LdG} = \frac{A_0}{2}\left(1 - \frac{U}{3}\right)\text{Tr}(\mathbf{Q}^2) - \frac{A_0 U}{3}\text{Tr}(\mathbf{Q}^3) + \frac{A_0 U}{4}\left(\text{Tr}(\mathbf{Q}^2)\right)^2 + \frac{L}{2}(\nabla \mathbf{Q})^2. \qquad (10)$$

The relaxation rate is controlled by the collective rotational diffusion constant $\Gamma$. The phenomenological coefficient $A_0$ sets the energy scale, $U$ controls the magnitude of the order parameter, and $L$ is the elastic constant in the one-elastic constant approximation. At the droplet surface with unit normal $\boldsymbol{\nu}$, the anchoring condition is imposed by adding a surface term to the free energy:

$$\mathcal{F}_{surf} = \int_{\partial V} f_{surf} dS, \qquad (11)$$

where the integration is performed over the surface of the droplet $\partial V$. Note that hybrid types of anchoring of the director field is enforced at the droplet surface. In a portion of the droplet with the homeotropic anchoring, the Rapini-Papoular like surface free energy density of the form

$$f_{surf} = \frac{1}{2}W(\mathbf{Q} - \mathbf{Q}^0)^2, \qquad (12)$$

is used which quadratically penalizes the deviation from the surface-preferred tensorial order parameter $\mathbf{Q}^0 = S_{eq}(\boldsymbol{\nu}\boldsymbol{\nu} - \mathbf{I}/3)$. The fourth order Fournier-Galatola free energy density is adopted to apply the degenerate planar anchoring boundary condition on the rest of droplet surface

$$f_{surf} = \frac{1}{2}W(\bar{\mathbf{Q}} - \bar{\mathbf{Q}}_\perp)^2 + \frac{1}{4}W(\bar{\mathbf{Q}} : \bar{\mathbf{Q}} - S_{eq}^2)^2, \qquad (13)$$

where $W$ controls the anchoring strength, $\bar{\mathbf{Q}} = \mathbf{Q} + \frac{1}{3}S_{eq}\boldsymbol{\delta}$, and its projection to the surface $\bar{\mathbf{Q}}_\perp = \mathbf{p} \cdot \bar{\mathbf{Q}} \cdot \mathbf{p}$ and $\mathbf{p} = \boldsymbol{\delta} - \boldsymbol{\nu}\boldsymbol{\nu}$. Assuming all of the transitions are axisymmetric (i.e. rotationally symmetric with respect to the director far field), patchy boundary conditions with different regions having either tangential or homeotropic anchoring were imposed on the droplet surface. While mimicking experimental observations, these boundary conditions were gradually changed axisymmetrically by advancing the location of the quarter-integer disclination along the polar-angle-changing direction.

The computer simulations were performed on a cubic mesh of $240 \times 240 \times 240$ uniformly distributed points. The total free energy of the system was minimized, corresponding to the nulls of first variation of $\mathcal{F}$, and the resultant Euler-Lagrange partial differential equations were solved using an explicit Euler relaxation finite difference scheme (*45-51*). The droplet radius was chosen to be $r_0 = 50.25$ in simulation unit. Formation of the hyperbolic hedgehog point defect was ensured by initializing the director field with the ansatz of the following form (*46*):

$$\mathbf{n} = \mathbf{n}_0 - Pr_0^2 \frac{\mathbf{d} - \mathbf{d}_0}{|\mathbf{d} - \mathbf{d}_0|^3}. \qquad (14)$$

where constant $P = 1.5$ determines the separation distance of the defect and the droplet center, $\mathbf{d}$ and $\mathbf{d}_0$ are the position vector and the position of the center of mass of the droplet, respectively.

The optical polarization micrograph textures were simulated using the method of Jones matrices and experimental material parameters (*52, 53*). In all of the simulated textures, the incident polarized wavevector propagated along the *y*-axis (perpendicular to the image), while polarizer and analyzer were crossed with the polarizer aligned along the far-field director. The following numerical parameters have been used throughout the simulations: $A_0 = 6\times10^5$ J m$^{-3}$, $U = 3.0$ yielding equilibrium value of $S_{eq} = 0.5$, $L = 6$ pN, and the anchoring strength in the range of $W = 6\times10^{-5}$ - $6\times10^{-3}$ J m$^{-2}$.

**Supplementary Materials**

Fig. S1. Composition of droplets.
Fig. S2. Comparison of the elastic hexadecapoles with mixed planar-homeotropic and conically degenerate surface anchoring.
Fig. S3. Elastic octupoles and their Brownian motion.
Movie S1. Optical microscopy textures during a transition from a boojum quadrupole to hedgehog dipole.
Movie S2. Computer simulations of a transition from a boojum quadrupole to Saturn ring quadrupole.
Movie S3. Computer simulations of a transition from a boojum quadrupole to octupole.
Movie S4. Computer simulations of a transition from a dipole to a boojum quadrupole.
Movie S5. Computer simulations of a director field change during a transformation of a defect loop into a boojum.

## Acknowledgments

**General**: We thank Q. Liu, B. Fleury and J.B. ten Hove for discussions. **Funding:** This work was partially supported by Department of Energy, Office of Basic Energy Sciences, under contract DE-SC0019293 with University of Colorado Boulder. K.C. acknowledges financial support from NSF and the Soft Materials Research Center (SMRC) Research Experience for Undergraduates (REU) program at the University of Colorado Boulder. **Author contributions:** B.S. and K.C. conducted the experiments. A.M. and J.J.d.P. developed the models and performed the simulations; R.Z. contributed to the theoretical analysis; I.I.S. conceived and directed the project. B.S. and I.I.S. analyzed the results and wrote the manuscript with input from all authors. **Competing interests:** The authors declare that they have no competing interests. **Data and materials availability:** All data needed to evaluate the conclusions in the paper are present in the paper and/or the supplementary materials. Additional data related to this paper may be requested from the authors.

## Figures and Tables

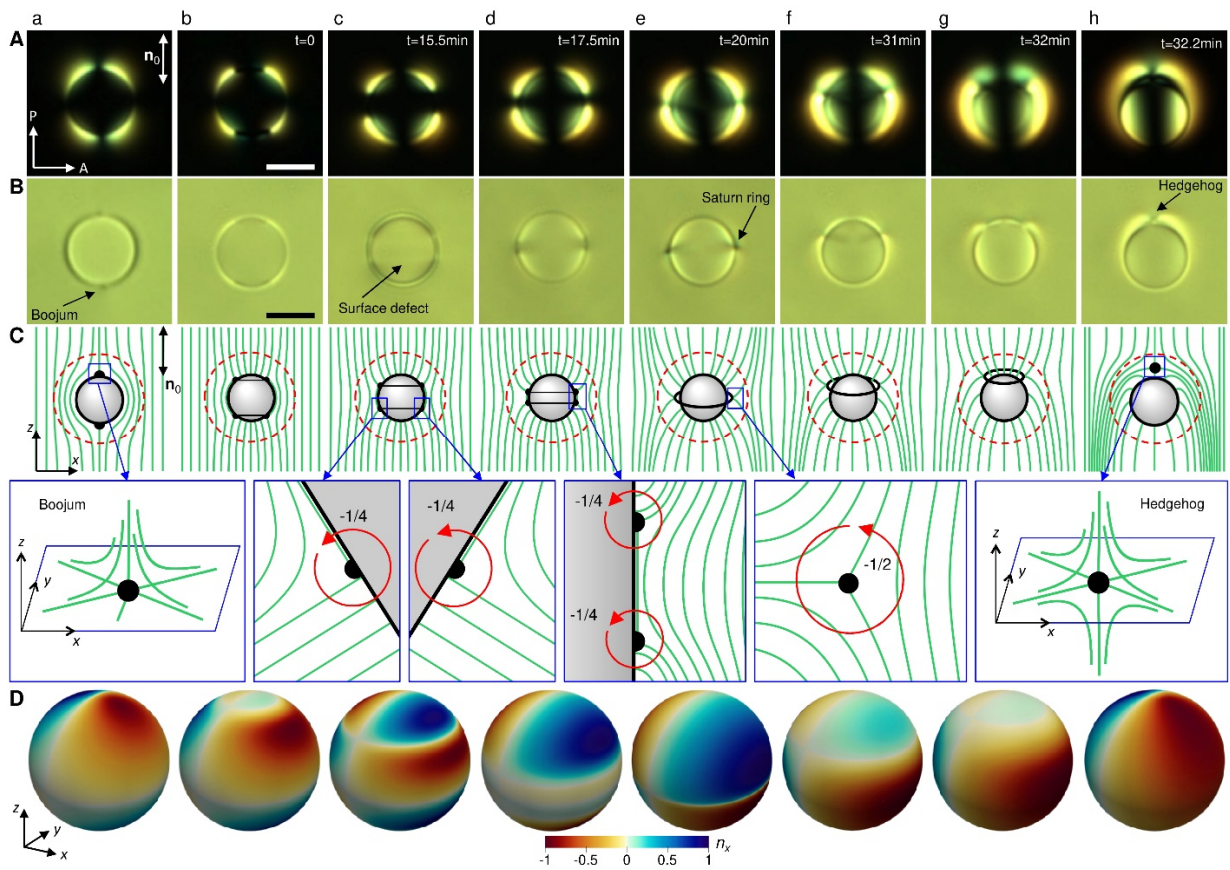

**Fig. 1. Transformations of elastic multipoles from a boojum quadrupole to a hedgehog dipole.** Polarizing (**A**) and brightfield (**B**) microscopy images of elastic multipoles during transformations; P and A mark crossed polarizer and analyzer and $\mathbf{n}_0$ is a far-field director. Schematics of the director field $\mathbf{n}(\mathbf{r})$ (**C**) and color maps of director distortions for elastic multipoles (**D**) corresponding to (**A**); color maps show the normalized $n_x$ component of the director on the shell enveloping the droplet and defects (red dashed circle in C). Scale bars, 5 μm. In the insets of (**C**), the winding numbers of quarter-integer and half-integer disclinations are marked.

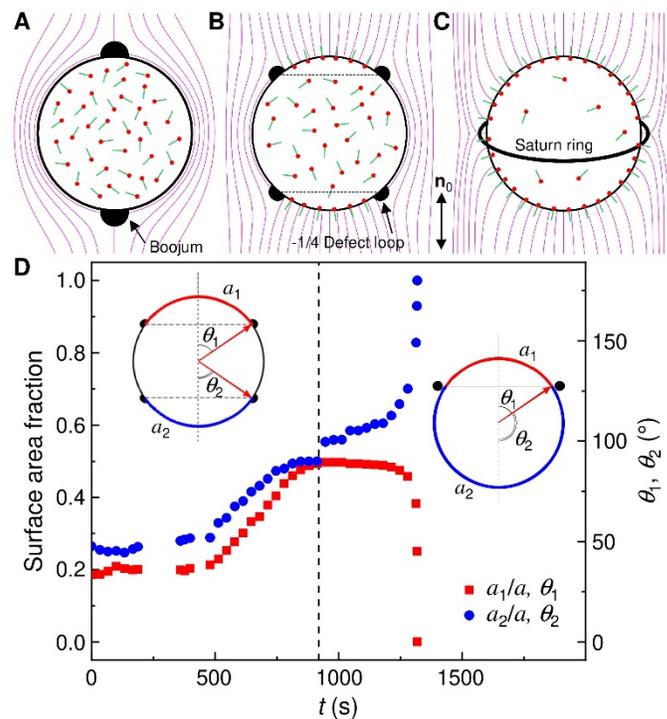

**Fig. 2. Anchoring transition at droplets surface and its dynamics.** (**A**-**C**) Schematics showing the mechanism of the anchoring transition at the interface between a droplet and LC. (**D**) Time dependence of an area coverage $a_1$, $a_2$ with homeotropic boundary conditions with respect to the total surface area $a$ of the droplet during the patchy anchoring transition and structural transformation.

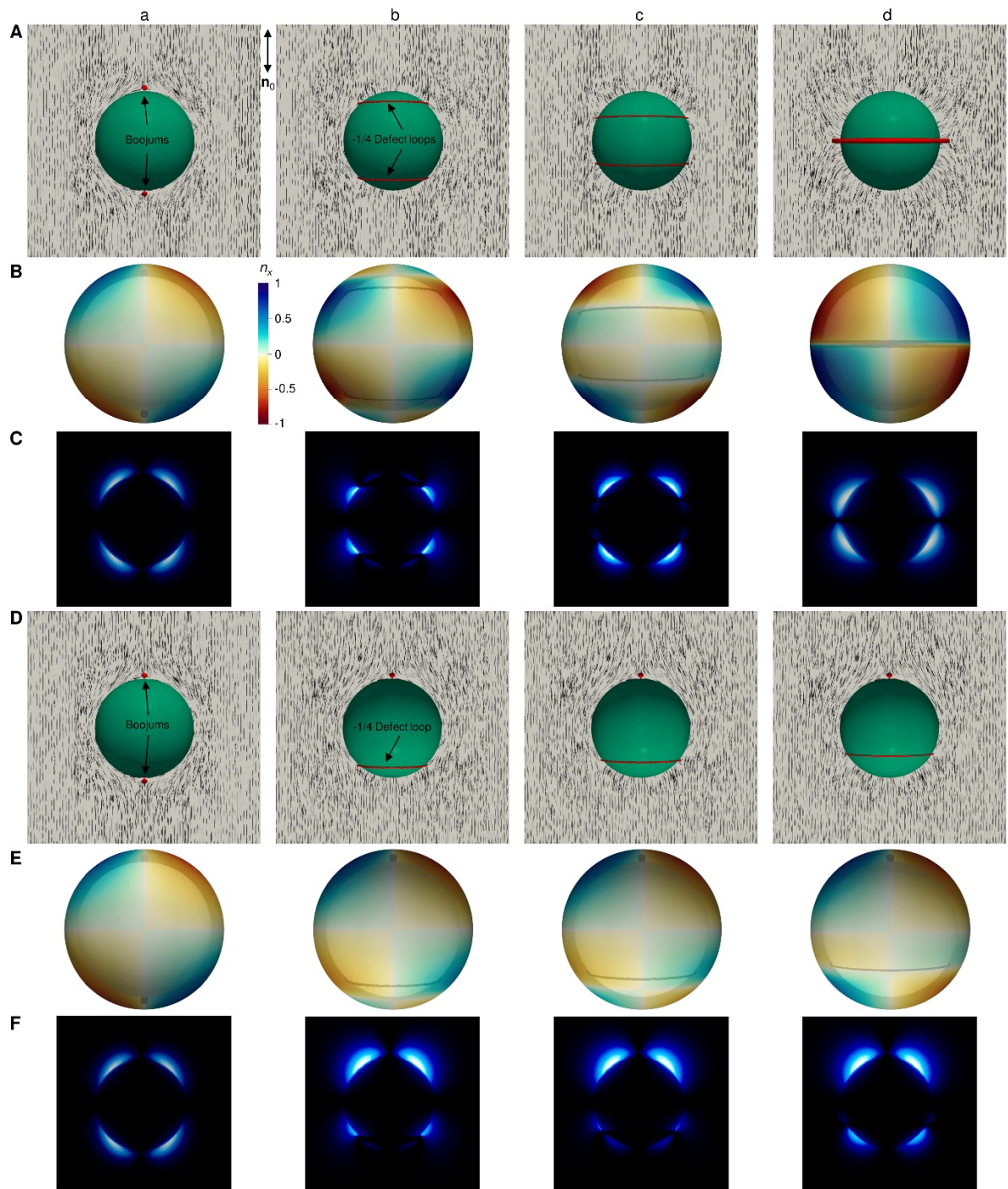

**Fig. 3. Computer simulations of director structures and elastic multipoles.** Simulated director field (**A**, **D**), elastic multipoles (**B**, **E**) and polarizing microscopy textures (**D**, **F**) for boojum - Saturn ring quadrupoles transformations (**A-C**) and boojum quadrupole to octupole transformations (**D-F**). The semi-transparent colored shells enveloping the droplet in **B** and **E** allow one to see the point and line defects in their interior, near the droplet surface. The radius of a mapping shell is $1.2r_0$. Red isosurfaces depict regions of a reduced scalar-order parameter $S = 0.3$. Crossed polarizer and analyzer in **D** and **F** are parallel respectively to vertical and horizontal sides of the image.

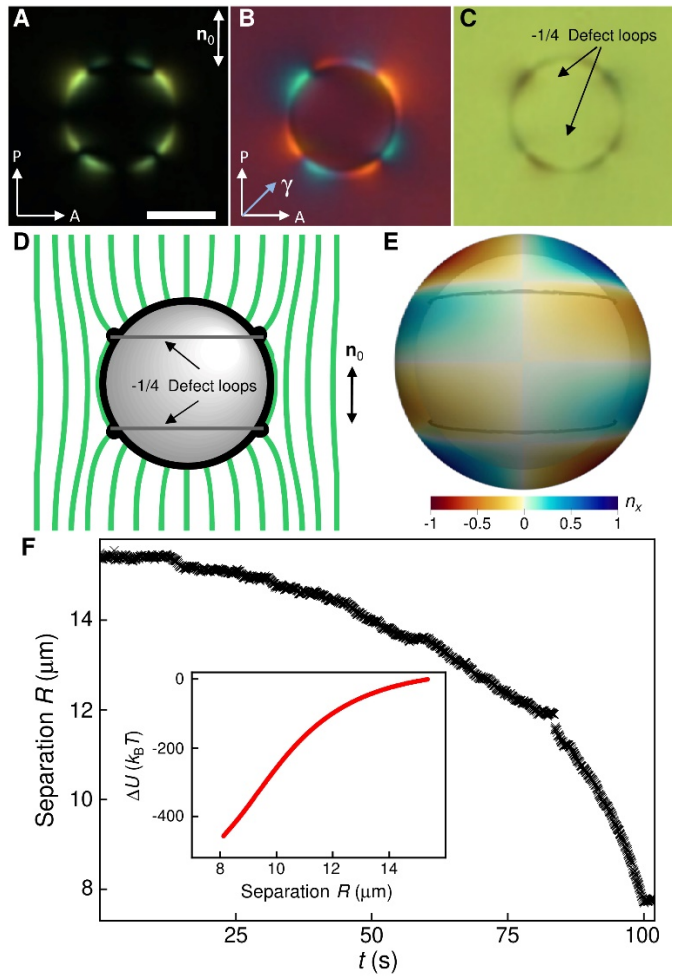

**Fig. 4. Elastic hexadecapoles with mixed surface anchoring.** (**A-C**) Optical microscopy textures of elastic hexadecapoles obtained between crossed polarizers without (**A**) and with (**B**) an inserted retardation plate and without polarizers (**C**); γ marks the direction of the retardation plate's slow axis. (**D**) Schematic diagram of the director field of hexadecapole and (**E**) a color map of the director's *x*-component ($n_x$) on the shell enclosing the droplet and defects. The radius of a droplet and a mapping shell in (**E**) is $1.2r_0$. (**F**) Separation versus time between two hexadecapoles with $(b_2,b_4,b_6)$=(0.083, 0.16, -0.0002) (from fitting with Eq. 5) attracting at $\theta \approx 38°$. The inset shows the corresponding interaction potential. Scale bar, 5 μm.

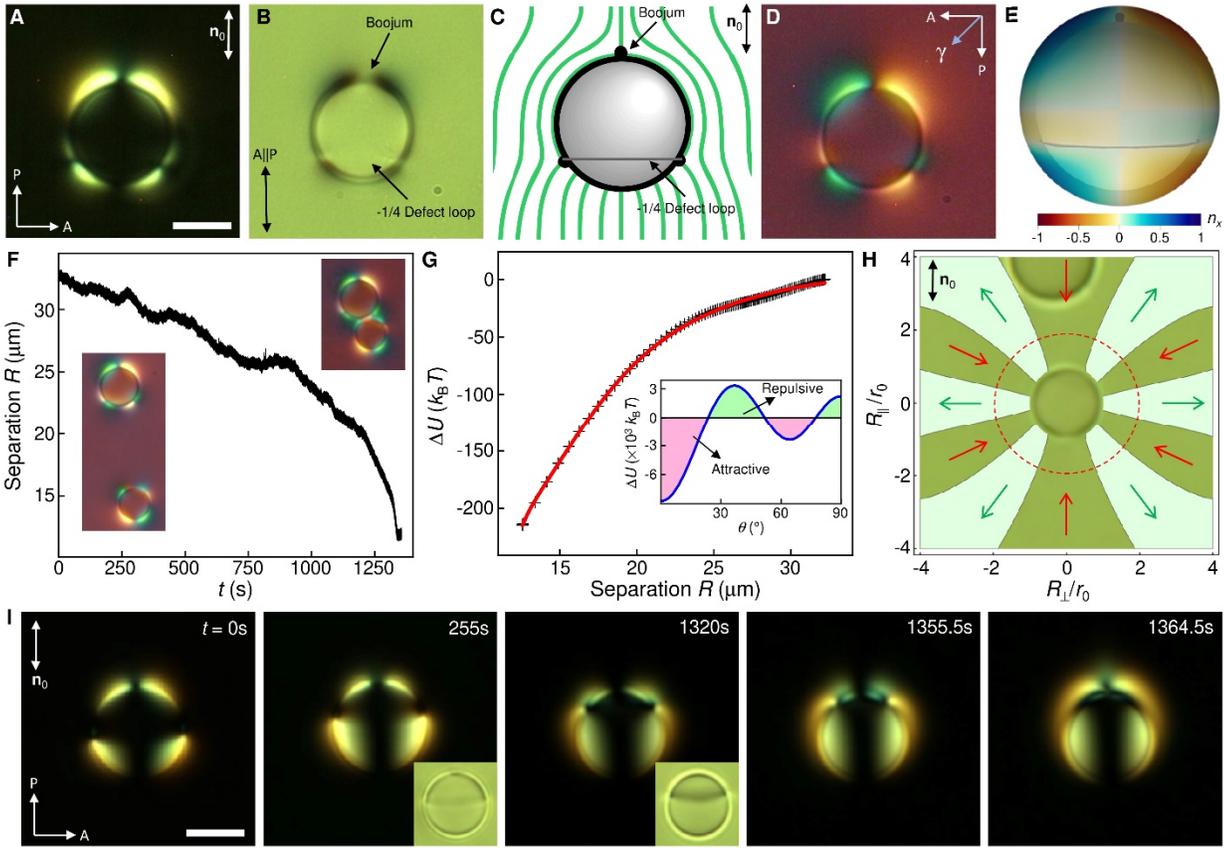

**Fig. 5. Elastic octupoles with mixed surface anchoring.** (**A,B,D**) Optical microscopy textures of elastic octupoles obtained between crossed polarizers without (**A**) and with (**D**) a red plate and without polarizers (**B**). (**C**) Schematic diagram of the director field of octupole and (**E**) a corresponding color map of the director's *x*-component ($n_x$) on the shell enclosing the droplet and defects. The radius of a mapping shell in (**E**) is $1.2r_0$. (**F**) Separation versus time between two octupoles attracting at $\theta \approx 23°$ and (**G**) corresponding interaction potential extracted using experimental data from (**F**). The insets in (**F**) show droplets at the beginning and in the end of interaction. The inset in (**G**) shows the angular dependence of pair-interactions of octupoles with $(b_1,b_3,b_5)=(0.182, 0.499, -0.005)$ obtained fitting experimental data with Eq. (5); magenta and green zones indicate attraction and repulsion, respectively. (**H**) Angular diagram showing zones of attraction and repulsion between octupoles calculated with elastic moments obtained by fitting data shown in (**G**): green arrows indicate repulsion and red arrows indicate attraction; the diagram is overlaid atop of a brightfield micrograph with two such droplets undergoing elastic interactions. $R_\parallel$ and $R_\perp$ are components of the center-to-center distance **R** respectively along and perpendicular to $\mathbf{n}_0$. (**I**) Sequence of micrographs showing the transformation of an elastic octupole to an elastic dipole with a hedgehog. Elapsed time is marked on images. Two representative brightfield micrographs are shown as insets. Scale bars, 5 μm.

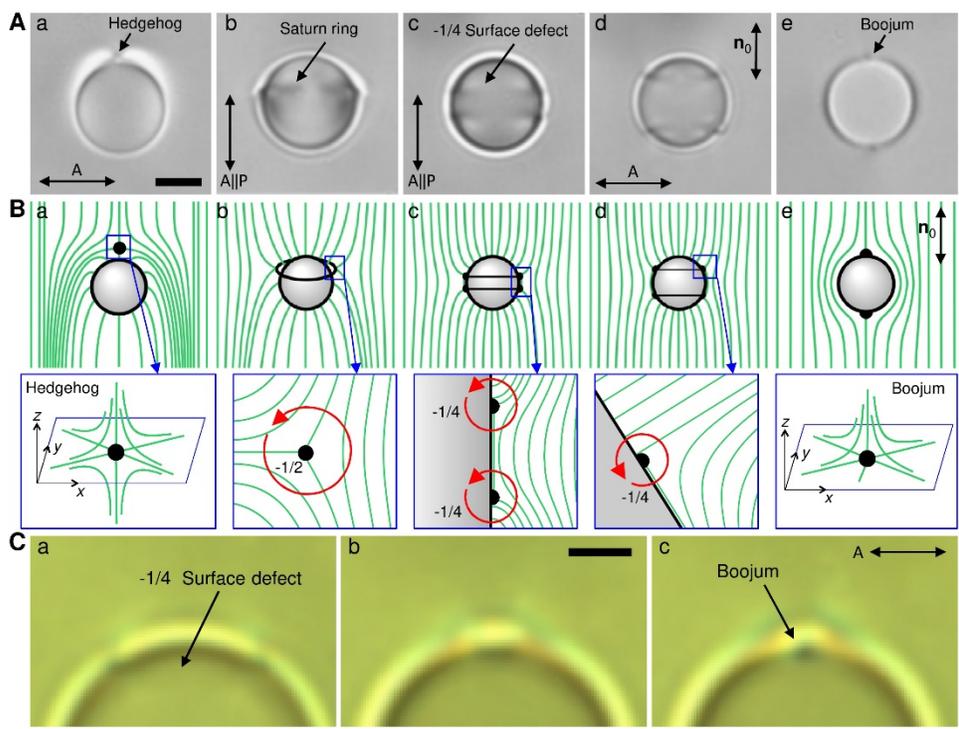

**Fig. 6. Hedgehog dipole to a boojum quadrupole transformation.** (**A**) Sequence of brightfield micrographs showing transformation of an elastic dipole into a boojum quadrupole. (**B**) Schematic diagrams of a director field structures and defects corresponding to (**A**); in the insets, the winding numbers of quarter-integer and half-integer disclinations are marked. (**C**) Sequence of zoomed-in brightfield micrographs showing details of the final-stage transformation of a surface defect loop into a point surface defect boojum. Scale bars, 5 μm.

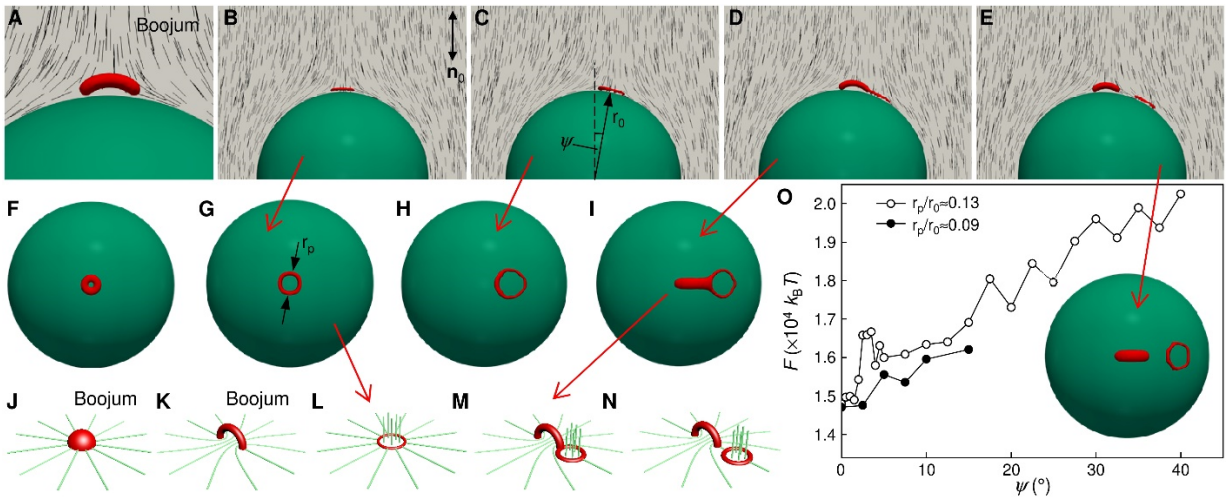

**Fig. 7. Topology and LC elasticity associated with patches of homeotropic anchoring.** (**A-E**) Side views of computer-simulated cross-sections containing the far-field LC director, showing structure of defects and director distortions of a droplet without surfactant self-assembly and with the boojum core in the form of handle-like half-integer disclination loop (**A**), when a self-assembled surfactant monolayer island nucleates at the pole and replaces the boojum with a closed loop of a quarter-integer surface disclination (**B**) and when it is away from the pole (**C**), as well as when the island is near the boojum (**D**) and away from it (**E**). (**F**, **G**) Top views of the droplet structures showing small (**F**) and larger (**G**) self-assembled surfactant islands circled by loops of quarter-integer disclinations when centered at the pole, like the one in (**B**). (**H**, **I**) Patches of homeotropic anchoring due to self-assembled surfactant islands when shifted away from the pole (**H**), corresponding to (**C**), and when nucleated away from the pole and fusing with a boojum defect at the pole, corresponding to (**D**). Ratios between radius of a patch with homeotropic anchoring $r_p$ and $r_0$ are $r_p/r_0 \approx 0.05$ in (**F**), $r_p/r_0 \approx 0.09$ in (**B**, **G**) and $r_p/r_0 \approx 0.13$ in (**C-E**, **H**, **I**, the inset of **O**). (**J-N**) Schematics of defect structure in the forms of a point surface boojum defect (**J**) and when it splits into a handle-like semi-loop of half-integer disclination (**K**) and when it is replaced with a quarter-integer closed disclination loop; the split-core boojum is shown fusing with the quarter-integer loop encircling a self-assembled island of surfactant (**M**) and when the boojum and disclination are separated by some distance (**N**). (**O**) Free energy associated with the droplet having a homeotropic patch surrounded by a quarter-strength disclination when shifted away from the pole or when nucleated away from the pole, demonstrating that it is always attracted to the pole of the drop. Note that the presence of the local barrier and local minimum in the dependence for a larger patch is associated with the energetic barrier associated with fusing of defect cores of the split-core boojum and quarter-integer disclination loop around the patch. The red regions in (**A-N**) correspond to the reduced scalar order parameter of $S = 0.3$.

# Supplementary Materials

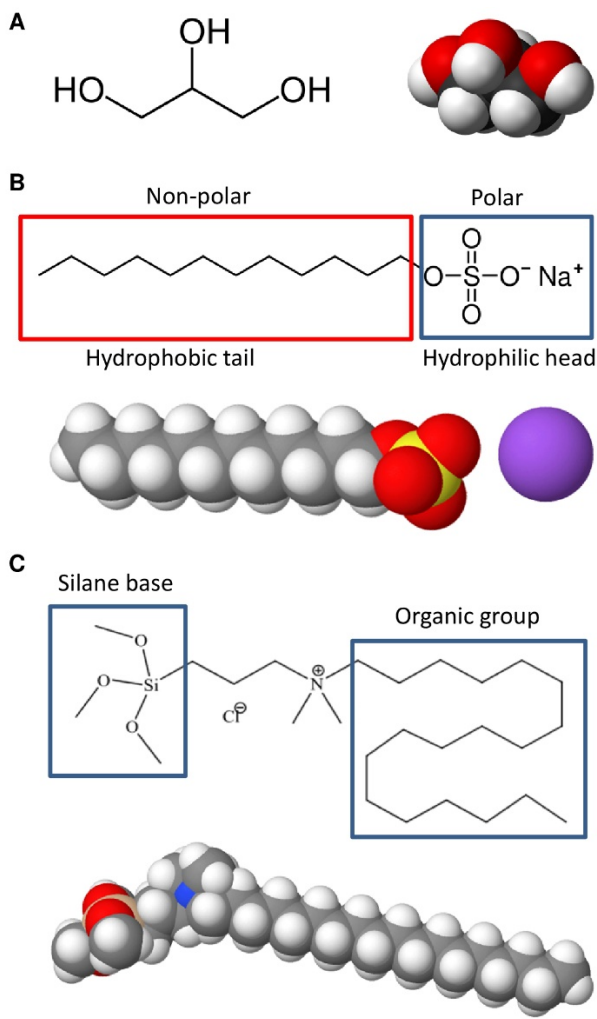

**Figure S1.** Composition of droplets: (A) Glycerol; (B) Sodium dodecyl sulfate (SDS); (C) N,N-dimethyl-N-octadecyl-3-aminopropyl-trimethoxysilyl chloride (DMOAP).

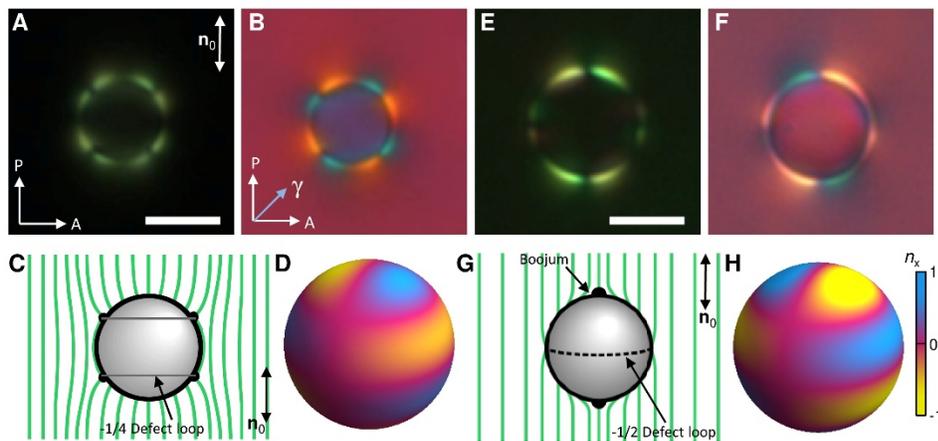

**Figure S2.** Comparison of the elastic hexadecapoles with (A-D) mixed planar-homeotropic and (E-H) conically degenerate surface anchoring. Scale bars, 5 μm.

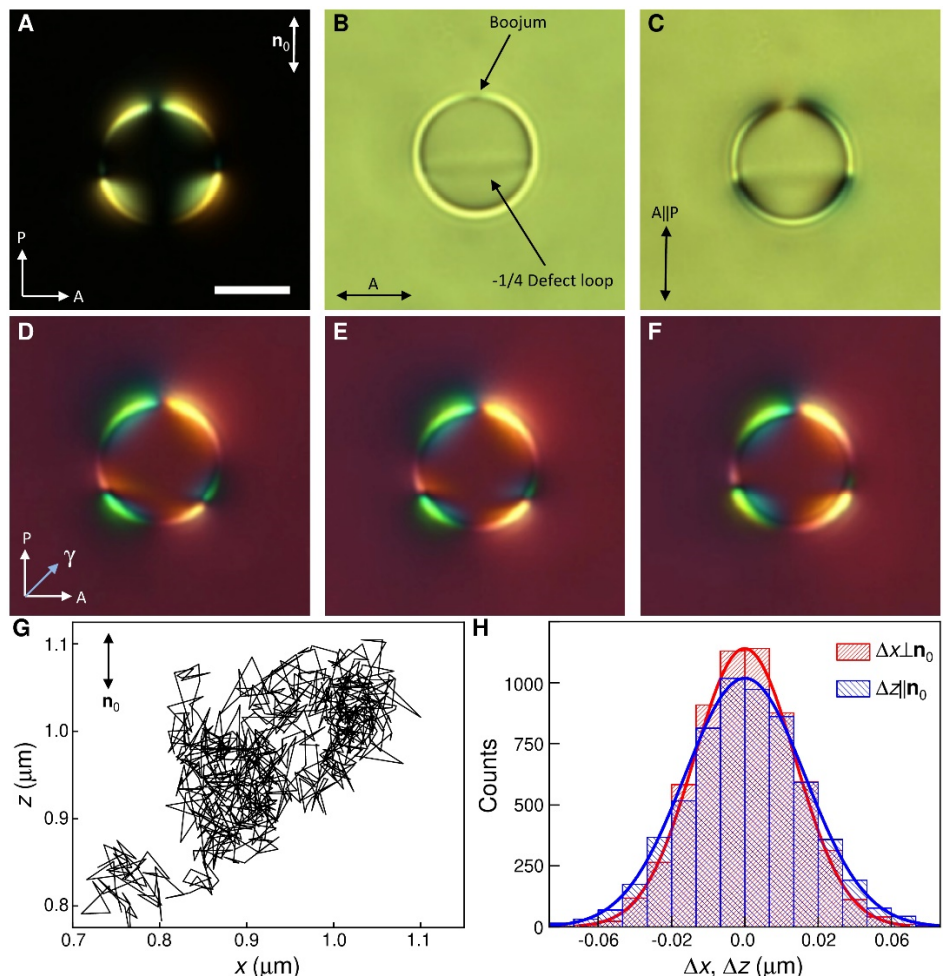

**Figure S3.** Elastic octupoles and their Brownian motion: (A-F) Optical micrographs between crossed polarizers (A, D-F) without (A) and with (D-F) a retardation plate and between parallel polarizers (C) and without a polarizer (B). (G) Trajectory of the octupole's Brownian motion obtained at a frame rate of 15 fps. (H) Histogram of octupole's displacements with respect to $\mathbf{n}_0$. Scale bars, 5 μm.

**Movie S1.** Optical microscopy textures during a transition from a boojum quadrupole to hedgehog dipole. Timer shows the elapsed time of transformations. P and A mark crossed polarizer and analyzer and $\mathbf{n}_0$ is a far-field director.

**Movie S2.** Computer simulations of a transition from a boojum quadrupole to Saturn ring quadrupole: (A) director field configurations with red regions showing the reduced scalar order parameter $S = 0.3$; (B) elastic multipoles shown as a color map of the director's $x$-component ($n_x$) on the shell enclosing the droplet and defects. The radius of a mapping shell is $1.2r_0$; (C) polarizing microscopy textures; P and A mark crossed polarizer and analyzer and $\mathbf{n}_0$ is a far-field director.

**Movie S3.** Computer simulations of a transition from a boojum quadrupole to octupole: (A) director field configurations with red regions showing the reduced scalar order parameter $S = 0.3$; (B) elastic multipoles shown as a color map of the director's $x$-component ($n_x$) on the shell enclosing the droplet and defects. The radius of a mapping shell is $1.2r_0$; (C) polarizing microscopy textures; P and A mark crossed polarizer and analyzer and $\mathbf{n}_0$ is a far-field director.

**Movie S4.** Computer simulations of a transition from a dipole to a boojum quadrupole: (A) director field configurations with red regions showing the reduced scalar order parameter $S = 0.3$; (B) elastic multipoles shown as a color map of the director's *x*-component ($n_x$) on the shell enclosing the droplet and defects; (C) polarizing microscopy textures; P and A mark crossed polarizer and analyzer and $\mathbf{n}_0$ is a far-field director.

**Movie S5.** Computer simulations of a director field change during a transformation of a defect loop into a boojum and the subsequent formation of a semi-ring of disclination corresponding to the boojum core. A red isosurface depicts a region with the reduced scalar order parameter $S = 0.3$.